# BOOTSTRAPPING CONFIDENCE LEVELS FOR HYPOTHESES ABOUT QUADRATIC (U-SHAPED) REGRESSION MODELS

12 June 2012

*Michael Wood*

University of Portsmouth Business School

SBS Department, Richmond Building

Portland Street, Portsmouth

PO1 3DE, UK

+44(0)23 9284 4168

michael.wood@port.ac.uk .

## Abstract

Bootstrapping can produce confidence levels for hypotheses about quadratic regression models – such as whether the U-shape is inverted, and the location of optima. The method has several advantages over conventional methods: it provides more, and clearer, information, and is flexible – it could easily be applied to a wide variety of different types of models. The utility of the method can be enhanced by formulating models with interpretable coefficients, such as the location and value of the optimum.

# BOOTSTRAPPING CONFIDENCE LEVELS FOR HYPOTHESES ABOUT QUADRATIC (U-SHAPED) REGRESSION MODELS

## Abstract

I show how bootstrapping can produce confidence levels for hypotheses about quadratic regression models – such as whether the U-shape is inverted, and the location of optima. The method has several advantages over conventional methods: it provides more, and clearer, information, and is flexible – it could easily be applied to a wide variety of different types of models. The utility of the method can be enhanced by formulating models with interpretable coefficients, such as the location and value of the optimum.

## 1. Introduction

Quadratic (U-shaped) models are widely used in management and economics – e.g. in research on well-being (Blanchflower and Oswald, 2008), staff turnover (Glebbeek and Bax, 2004) and the environmental Kuznets curve (Dinda, 2004). They are conventionally analyzed by reporting regression coefficients, and $p$ values, for the independent variable and its square. However, there is a very large literature on the problems with $p$ values in general terms (e.g. Nickerson, 2000), and they are particularly problematic for hypotheses about U-shapes: there are two $p$ values which makes it difficult to get a clear idea of the degree of support for the U-shape hypothesis.

This paper shows how simple bootstrapping can be used to derive confidence levels for hypotheses about quadratic regression models. The approach is intuitive, flexible, avoids the distributional assumptions required by conventional methods, and provides users with direct estimates of the (posterior) probabilities of the competing hypotheses. I also suggest reformulating the standard model in terms of more easily interpreted coefficients.

Most applications of bootstrapping use relatively complex, possibly novel, approaches to intractable problems. This paper differs in two respects: first, the bootstrap method used is a very simple one, and second, the problem tackled is widely seen as solved. The originality of this paper lies in showing how a very simple method can provide better answers to the problem than the standard ones.

Figure 1 shows one model discussed by Glebbeek and Bax (2004). (Glebbeek and Bax do not show this diagram. Dr. Glebbeek, however, was kind enough to give me access to the data.) They investigated the hypothesis that there is an "inverted U-shape relationship" between two variables – staff turnover and organizational performance. The hypothesis is based on the idea that both high and low levels of staff turnover are likely to lead to poor performance, so there will be an optimum level of turnover (e.g. 10% of the staff leaving in a year) with performance falling off above and below this. For an inverted U shape the regression coefficient for the squared turnover term should be negative, and for the linear term the coefficient should be positive. In the model shown in Figure 1 the signs of the coefficients were consistent with this hypothesis but neither was significant ($p$>0.1 in both cases).

**Figure 1. Predicted performance from quadratic model (after adjusting for values of three control variables)**

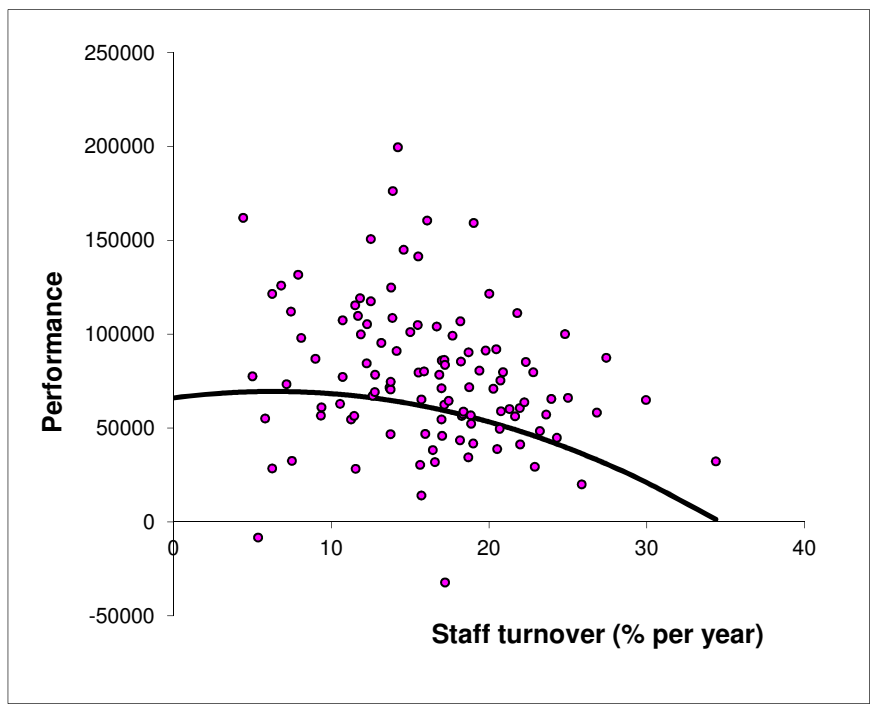

(The solid line is the prediction from the regression model; the scattered points are the data on which the regression model is based.)

## 2. Rewriting the model with easily interpreted coefficients

The equation of the curve is usually written as:

$$y = ax^2 + bx + c$$

where x is the independent variable (turnover), *y* is the dependent variable (performance), and *c* is constant (and includes the adjustment for the control variables). Alternatively, exactly the same curve can be written as

$$y = M + U(x - L)^2$$

The square term in this equation must be zero or positive, so *M* is the maximum value of *y* if *U*<0 and the minimum if *U*>0. *L* is obviously the location (*x* coordinate) of this maximum or minimum, and *U* is the degree of curvature with positive values corresponding to an upright U-shape and higher values (positively or negatively) indicating a greater degree of curvature. Comparing the two equations:

$$U = a, L = - b/2a, M = c - b^2/4a.$$

In Figure 1, *M* = 69575, *L* = 6.3% and *U* = -86.7. Each coefficient then has an obvious interpretation. Obviously, for an inverted U-shape, *L*>0 (since the turning point must occur for positive *x*) and *U*<0.

## 3. Bootstrapping confidence levels for hypotheses

The question now arises of whether this demonstrates that a similar pattern would occur if the analysis was done with the whole population from which the sample was drawn. Conventionally this question is answered by testing two null hypotheses – the first being that the coefficient of the linear term is zero, and the second being that the squared term is zero. In the model represented by Figure 1, neither coefficient is significantly different from zero. The evidence provides some support for the inverted U-shape hypothesis, but it is difficult to combine the two *p* values into a single figure to indicate the overall strength of the support for this hypothesis.

A bootstrap method provides both a way out of this difficulty, and an easy approach to other questions we are likely to have. The result from the bootstrap analysis below is that the data on which Figure 1 is based suggests a confidence level, or probability, of 67% for the inverted U shape hypothesis. The method is implemented on an Excel spreadsheet (available on the web) which can easily be adapted to analyze different datasets or models.

Bootstrapping involves taking a large number of resamples (say 1000) with replacement from a sample of data. (For each resample we choose a member of the original sample at random, then replace it and choose again, until we have a sample the same size as the real sample. This means that some members of the original sample will appear in the resample more than once, and others not at all.) The simplest approach – which I use here – is then simply to assume that each resample represents one possible distribution for the original population, and that the collection of 1000 resamples represents a probability distribution for these possibilities. I will consider the justification for this assumption after I have explained how it works in our example.

**Figure 2. Predicted performance using a quadratic model (after adjusting for values of three control variables) – from the data (bold) and three resamples (dotted lines)**

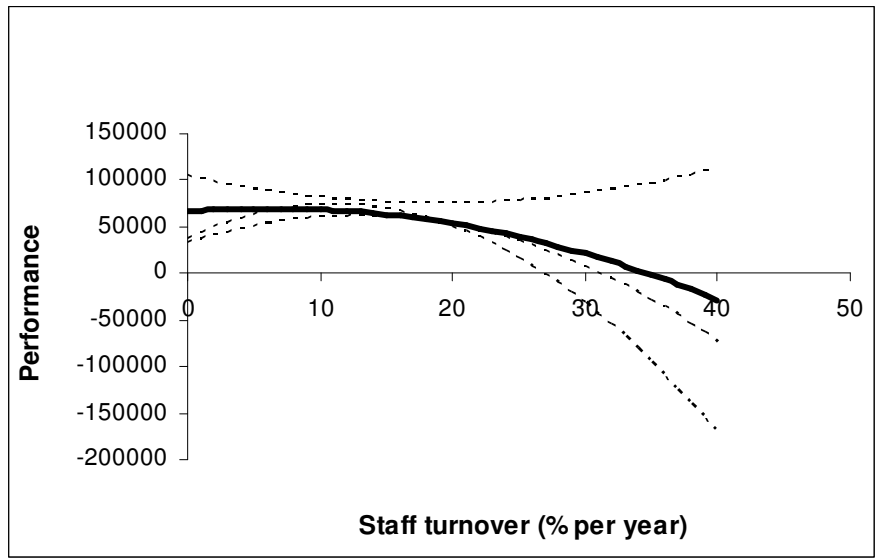

Figure 2 shows results from three resamples as well as the original sample. Each of the dotted lines in this figure is based on identical formulae to the solid line representing the real sample, but using the data from a resample, rather than the original sample. Two of these resamples are obviously an inverted U shape; the third is not.

These results come from an Excel spreadsheet at
http://woodm.myweb.port.ac.uk/BRQ.xls. The Resample sheet of the spreadsheet allows users
to press the Recalculate button (F9) and generate a new resample and line on the graph.

It is then a simple matter to produce more of these resamples and count up how many
of them are inverted U-shapes. The conclusion was that 67% of 1000 resamples gave an
inverted U shape (with $L$>0), which suggests that the confidence level for this hypothesis, based
on the data, should be put at 67%. (The same method and spreadsheet can also be used to
produce confidence bands on the graph.)

## 4. Assumptions and interpretation

The assumption made above that each resample represents a different possible population from
which the original sample was drawn is just an assumption. Resampling with replacement
simulates the process of drawing samples from an infinite population with the same distribution
as the original sample, so the resamples represent different *samples* that might have been
drawn from this *population*. For example, the top resample in Figure 2 – the one which is a non-
inverted U-shape – shows that even if the overall population pattern shows an inverted U-shape
this may not be true of individual samples. The discrepancies between the dotted lines and the
bold lines in Figure 2 illustrate the extent of sampling error, so from this point of view it seems
reasonable to assume that we can reverse this argument and assume that the dotted lines
represent different *populations* from which the *sample* might have been drawn.

There is a large literature on testing various elaborations on the basic bootstrap method
in various different situations, with conclusions that are predictably mixed. However, experience
shows that in "well-behaved" situations the bootstrap approach gives similar results for
confidence intervals to standard methods based on probability theory. For example, 1000
bootstrap resamples gives a 95% confidence interval for the slope of the linear model (derived
from the data in Figure 1) of –3147 to –704; the corresponding estimate from the formulae
based on the $t$ distribution built into Excel is –3060 to –495. In this case the bootstrap interval is
5% narrower (and the next bootstrap interval was 3% wider).

Furthermore, confidence intervals derived by conventional, frequentist methods are
often identical to Bayesian credible intervals based on flat prior distributions (Bayarri and Berger, 2004). This suggests that it is reasonable to interpret bootstrapped confidence levels in probabilistic terms as Bayesian posterior probabilities, making the assumption that the prior distribution is uniform in the sense that each of resample results, like the four shown in Figure 2, are equally likely.

## 5. Confidence levels for other hypotheses

It is very easy to use this method to obtain confidence levels for other hypotheses. The proportion of resamples exhibiting *any* specified property can easily be worked out. For example, we might want to know the location ($L$) of the optimum staff turnover. The point estimate from the regression shown in Figure 1 is that the optimum performance occurs with a staff turnover of 6.3%. The 1000 resamples give these confidence levels for three hypotheses:

- Confidence in hypothesis that $L$ is between 0% and 10%        = 30%

- Confidence in hypothesis that $L$ is between 10% and 20%      = 37%

- Confidence in hypothesis that $L$ is above 20%                      = 0%

Alternatively we might decide that an inverted U-shape needs a minimum difference between the optimum performance ($M$) and the value corresponding to a turnover level of zero. Setting this minimum to 10,000 units, the confidence level then comes to 40%.

Another hypothesis of interest to Glebbeek and Bax (2004) is that the relationship between performance and turnover is negative – this being the rival to the inverted U shape hypothesis. The top resample in Figure 2 illustrates the importance of defining this clearly: this shows a negative relationship for low turnover, but a positive relation for higher turnover. If we define a negative relationship as one which is not an inverted U shape, and for which the predicted performance for Turnover = 25% is less than the prediction for Turnover = 0%, then the spreadsheet shows that

- Confidence in hypothesis that the relationship is negative                  = 33%

In fact, all 1000 resamples gave either an inverted U shape or a negative relationship in this sense.

This result gives a very different picture from a standard linear regression analysis. The linear regression coefficient is negative with a $p$ value of 0.007: this is equivalent to a confidence level for a negative relationship of 99.6%. The reason for the difference is that each method has a different underlying model (quadratic in the first case, linear in the second) and defines a negative relationship differently: the bootstrap method has the advantage that it can be used with *any* definition of a negative relationship.

## 6. Conclusions

I have shown how bootstrapping can be used to estimate confidence levels for whatever hypotheses are of interest. In the example, instead of two $p$ values which are difficult to interpret, the conclusion is that we can be 67% confident about the inverted U-shape hypothesis, and 37% confident that the optimum turnover is more than 10%. Furthermore, the bootstrap method provides an intuitive idea of sampling variability, and the results can be interpreted as probabilities. The transparency of the approach can also be enhanced by reformulating models in terms of the location and value of the optimum, and the degree of curvature (instead of the conventional coefficients for the linear and square terms).

This is just one example. Quadratic models are widely used, and similar methods could easily be used to analyze hypotheses about other models.

## Acknowledgment

I am very grateful to Dr. Arie Glebbeek for making his data available to me.